\documentclass[floats, prd, eqnum, showpacs, nofootinbib, twocolumn,eqsecnum]{revtex4-1}

\usepackage{color,graphicx}
\usepackage{amsfonts}
\usepackage{amssymb}
\usepackage{comment}
\begin{document}

\title{Retrolensing by light rays slightly inside and outside of a photon sphere around a Reissner-Nordstr\"{o}m naked singularity} 
\author{Naoki Tsukamoto${}^{1}$}\email{tsukamoto@rikkyo.ac.jp}

\affiliation{
${}^{1}$Department of General Science and Education, National Institute of Technology, Hachinohe College, Aomori 039-1192, Japan \\
}

\begin{abstract}
We investigate the retrolensing of sunlights reflected by a photon sphere and by a potential barrier near an antiphoton sphere 
around a Reissner-Nordstr\"{o}m naked singularity.
We apply the deflection angles of the light rays in strong deflection limits to the retrolensing. 
We show that the retrolensing by the photon sphere around the Reissner-Nordstr\"{o}m naked singularity can be brighter 
than the one around a Reissner-Nordstr\"{o}m black hole because of the rays reflected by the potential barrier.
\end{abstract}
\maketitle

\section{Introduction}
Gravitational lensing with a small deflection angle $\alpha \ll 1$ 
has been investigated eagerly for finding dark and massive objects 
such as dark matters and extrasolar planets~\cite{Schneider_Ehlers_Falco_1992}.
On the other hand, the images of light rays with a large deflection angle $\alpha > 1$ 
in a strong gravitational field have been considered intermittently. 
In 1931~\cite{Hagihara_1931}, Hagihara pointed out that an observer can see light rays reflected by a photon sphere, 
which is a sphere made of unstable circular light orbits, 
and Darwin~\cite{Darwin_1959} and several authors revisited the images~\cite{Atkinson_1965,Luminet_1979,Ohanian_1987,
Nemiroff_1993,Virbhadra:1998dy,Frittelli_Kling_Newman_2000,Virbhadra_Ellis_2000,Bozza_Capozziello_Iovane_Scarpetta_2001,Bozza:2002zj,Virbhadra:2002ju,
Perlick:2003vg,Virbhadra:2008ws,Bozza_2010,Tsukamoto:2016zdu,Shaikh:2019jfr,Shaikh:2019itn,Tsukamoto:2020uay,Tsukamoto:2020iez,Paul:2020ufc,Perlick_2004_Living_Rev,Claudel:2000yi,Perlick:2021aok}.
Phenomena related to the photon sphere are more significant than before;
gravitational waves emitted by black hole binaries have been reported by LIGO and Virgo Collaborations~\cite{Abbott:2016blz}
and a shadow at the center of a giant elliptical galaxy M87 has been detected by the Event Horizon Telescope Collaboration~\cite{Akiyama:2019cqa}.
Theoretical and observational aspects of the photon sphere and an antiphoton sphere which which is a sphere made of stable circular light orbits~\footnote{It 
is of concern that the antiphoton sphere may lead to instability of compact objects 
because of the slow decay of linear waves~\cite{Keir:2014oka,Cardoso:2014sna,Cunha:2017qtt}.}
have been investigated~\cite{Hod:2017xkz,Sanchez:1977si,Press:1971wr,Abramowicz:1990cb,Koga:2016jjq,Barcelo:2000ta,Ames_1968}.
The generalizations or alternative surfaces of the photon sphere have been also suggested by several researchers~\cite{Claudel:2000yi,Gibbons:2016isj}.

In 2002, Bozza~\cite{Bozza:2002zj} investigated the gravitational lensing of the light rays reflected by the photon sphere 
in a general asymptotically-flat, static, and spherical symmetric spacetime in a strong deflection limit $b \rightarrow b_\mathrm{m}+0$,
where $b$ is the impact parameter of the light and $b_\mathrm{m}$ is its critical impact parameter.
The deflection angle of the light ray is expressed by 
\begin{eqnarray}\label{eq:al1}
\alpha &=&- \bar{a} \log \left( \frac{b}{b_\mathrm{m}}-1 \right) +\bar{b} \nonumber\\
&&+O \left( \left( \frac{b}{b_\mathrm{m}}-1 \right) \log \left( \frac{b}{b_\mathrm{m}}-1 \right) \right),
\end{eqnarray}
where $\bar{a}$ and $\bar{b}$ are determined by the metric of the spacetime.~\footnote{In Ref.~\cite{Bozza:2002zj}, the order of the vanishing term 
in the deflection angle in the strong deflection limit is considered as $O(b-b_\mathrm{m})$ 
but we should read the order as $O \left( \left( b/b_\mathrm{m}-1 \right) \log \left( b/b_\mathrm{m}-1 \right) \right)$ 
as shown in Refs.~\cite{Iyer:2006cn,Tsukamoto:2016qro,Tsukamoto:2016jzh}.
}
The gravitational lensing in the strong deflection limit has been investigated in many aspects, 
and the details of its analysis have been studied~\cite{Eiroa:2002mk,Eiroa:2003jf,Bozza:2004kq,Bozza:2005tg,Bozza:2006sn,Bozza:2006nm,Iyer:2006cn,Bozza:2007gt,Tsukamoto:2016qro,Ishihara:2016sfv,Tsukamoto:2016oca,Tsukamoto:2016zdu,Tsukamoto:2016jzh,Tsukamoto:2017edq,Shaikh:2019jfr,Shaikh:2019itn,Tsukamoto:2020uay,Tsukamoto:2020iez,Paul:2020ufc,Hsieh:2021scb,Aldi:2016ntn,Takizawa:2021gdp,Tsukamoto:2020bjm,Tsukamoto:2021caq,Aratore:2021usi}.

In 2019, Shaikh \textit{et al.}~\cite{Shaikh:2019itn} investigated the gravitational lensing of rays which pass through a photon sphere 
and which are reflected by a potential barrier near an antiphoton sphere in 
a general asymptotically-flat, static and spherical symmetric spacetime without an event horizon.
The deflection angle in a strong deflection limit $b \rightarrow b_\mathrm{m}-0$ is expressed by 
\begin{eqnarray}\label{eq:al2}
\alpha &=&- \bar{c} \log \left( \frac{b_\mathrm{m}^2}{b^2}-1 \right) +\bar{d} \nonumber\\
&&+O \left( \left( \frac{b_\mathrm{m}}{b}-1 \right) \log \left( \frac{b_\mathrm{m}}{b}-1 \right) \right),
\end{eqnarray}
where $\bar{c}$ and $\bar{d}$ are calculated 
from the metric of the spacetime.~\footnote{
We can approximate $b_\mathrm{m}^2/b^2-1 \sim 2\left( b_\mathrm{m}/b-1 \right) \sim 2\left( 1-b/b_\mathrm{m} \right)$, 
but we keep the form of Eq. (\ref{eq:al2}) as well as Ref.~\cite{Shaikh:2019itn}.}

Compact objects in nature would not have a large amount of an electric charge 
since the charged compact objects are neutralized quickly. 
However, in general relativity, the Reissner-Nordstr\"{o}m spacetime 
is often considered as a toy model of a compact object since
its property would be similar to other spacetimes, such as a Hayward spacetime~\cite{Chiba:2017nml},
and since we might treat it analytically.  
A shadow~\cite{deVries:2000,Takahashi:2005hy,Zakharov:2014lqa,Akiyama:2019cqa,Akiyama:2019eap,Kocherlakota:2021dcv}, time delay of rays~\cite{Sereno:2003nd},
and gravitational lensing~\cite{Bin-Nun:2010exl,Bin-Nun:2010lws} in the Reissner-Nordstr\"{o}m spacetime 
have been investigated.

Eiroa \textit{et al.} have considered the deflection angle in the strong defection limit $r_0\rightarrow r_\mathrm{m}+0$, where $r_0$ and $r_m$
are the radial coordinate of the closest position of a light ray and the position of the photon sphere, respectively, or 
$b\rightarrow b_\mathrm{m}+0$ numerically 
in the Reissner-Nordstr\"{o}m spacetime~\cite{Eiroa:2002mk}.
Bozza~\cite{Bozza:2002zj} obtained the analytical form of $\bar{a}$ in Eq.~(\ref{eq:al1}) 
and $\bar{b}$ was partly calculated numerically in the strong deflection limit $b \rightarrow b_\mathrm{m}+0$.
The exact analytical expressions of $\bar{a}$ and $\bar{b}$ have been obtained by Tsukamoto and Gong~\cite{Tsukamoto:2016jzh} 
and by Tsukamoto~\cite{Tsukamoto:2016oca}.
The gravitational lensing in the strong deflection limit, $b\rightarrow b_\mathrm{m}+0$, by a marginally unstable photon sphere 
has also been considered~\cite{Tsukamoto:2020iez}.
Gravitational lensing by a photon sphere around a naked singularity in the Reissner-Nordstr\"{o}m spacetime with $q^2/m^2=1.05$ 
in the strong defection limit $b\rightarrow b_\mathrm{m}-0$ has been investigated by Shaikh \textit{et al.}~\cite{Shaikh:2019itn}
and exact analytical forms of $\bar{c}$ and $\bar{d}$ in Eq.~(\ref{eq:al2}) have been calculated by Tsukamoto~\cite{Tsukamoto:2021fsz}.

Holz and Wheeler~\cite{Holz:2002uf} have investigated retrolensing with the deflection angle $\alpha \sim \pi$ in a lens configuration that 
has the Sun as the source of a light, an observer, and a black hole as a lens object line up in this order.
Retrolensing by a photon sphere around a black hole~\cite{DePaolis:2003ad,Eiroa:2003jf,Bozza:2004kq,DePaolis:2004xe,Abdujabbarov:2017pfw},
around a wormhole~\cite{Tsukamoto:2017edq,Tsukamoto:2016zdu},
and a naked singularity~\cite{ZamanBabar:2021zuk} have been investigated.
Retrolensing with a deflection angle $\alpha \sim 3\pi$ also have been studied~\cite{Tsukamoto:2017edq}.
Retrolensing by a Reissner-Nordstr\"{o}m black hole in our galaxy~\cite{Eiroa:2003jf} 
and near our solar system~\cite{Tsukamoto:2016oca} have been studied.

In this paper, we apply the retrolensing of light rays reflected near a photon sphere 
around a Reissner-Nordstr\"{o}m naked singularity
by using the exact analytic expressions of $\bar{a}$ and $\bar{b}$ in Eq.~(\ref{eq:al1}) 
in the strong deflection limit $b \rightarrow b_\mathrm{m}+0$ 
and $\bar{c}$ and $\bar{d}$ in Eq.~(\ref{eq:al2}) in the strong deflection limit $b \rightarrow b_\mathrm{m}-0$. 
The retrolensing by the photon sphere around the Reissner-Nordstr\"{o}m naked singularity can be brighter 
than the one around a Reissner-Nordstr\"{o}m black hole 
because of the rays reflected by a potential barrier near an antiphoton sphere.

This paper is organized as follows. 
We investigate the retrolensing near the photon sphere around the Reissner-Nordstr\"{o}m naked singularity in Sec.~II,
and we discuss and conclude our results in Sec.~III.
We review the deflection angle of rays in the strong deflection limits in the Reissner-Nordstr\"{o}m spacetime briefly in Appendix A. 
In this paper, we use the units in which the light speed and Newton's constant are unity.

\section{Retrolensing in the Reissner-Nordstr\"{o}m spacetime}
\begin{figure}[htbp]
\begin{center}
\includegraphics[width=70mm]{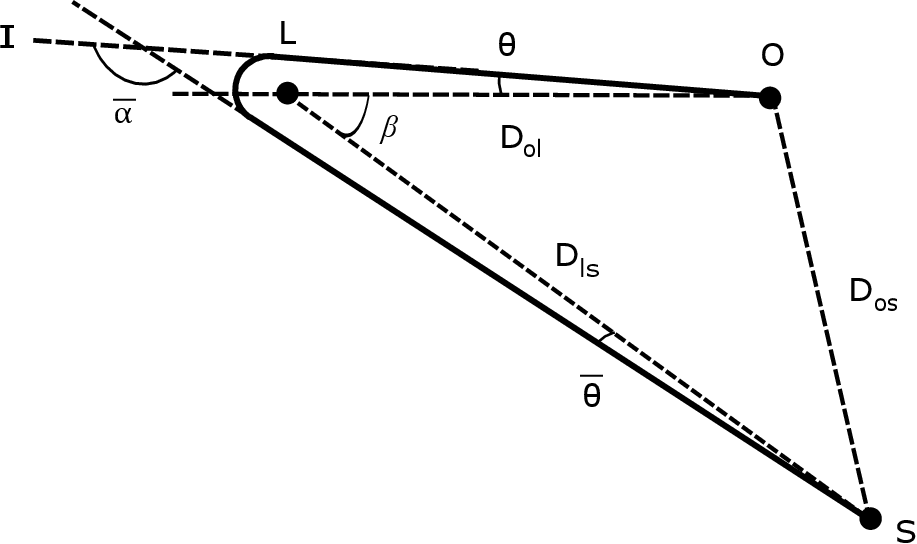}
\end{center}
\caption{Lens Configuration.
The ray of the Sun S at the source angle $\beta\equiv \angle$OLS is reflected by a photon sphere L with an effective deflection angle $\bar{\alpha}$. 
An angle $\theta$ denotes the image angle of image I observed by an observer O 
and $\bar{\theta}$ denotes an angle between the line~LS and the ray at S.
}
\label{Lens_Configuration}
\end{figure}

In this section we investigate the retrolensing of rays outside and inside of the photon sphere, 
the percent errors of the deflection angles in the strong deflection limits, and retrolensing light curves
in the Reissner-Nordstr\"{o}m spacetime.
As shown in Fig.~\ref{Lens_Configuration}, we consider a lens configuration where the ray is emitted by the Sun S, 
it is reflected by the photon sphere as a lens L 
with a deflection angle $\alpha$, and it reaches an observer O.
An image I with an image angle $\theta$ can be seen by the observer O.
We define an effective deflection angle $\bar{\alpha}$ as 
\begin{equation}
\bar{\alpha} \equiv \alpha -2\pi n,
\end{equation}
where $n$ is the winding number of the ray.

By using a source angle $\beta \equiv \angle$OLS defined in the domain 0~$\leq \beta \leq \pi$
and an angle $\bar{\theta}$ between the line~LS and the light ray at~S,
the Ohanian lens equation~\cite{Ohanian_1987,Bozza:2004kq,Bozza:2008ev} is expressed by
\begin{equation}\label{eq:Lens1}
\beta=\pi-\bar{\alpha}(\theta)+\theta+\bar{\theta}.
\end{equation}
We assume that the photon sphere~L, the observer~O, and the Sun~S 
are almost aligned in this order. This yields
\begin{equation}
\beta \sim 0,
\end{equation}
\begin{equation}
\bar{\alpha} \sim \pi,
\end{equation}
\begin{equation}
\alpha \sim \pi +2\pi n,
\end{equation}
and
\begin{equation}
D_\mathrm{ls}=D_\mathrm{ol}+D_\mathrm{os},
\end{equation}
where $D_\mathrm{ls}$, $D_\mathrm{ol}$, and $D_\mathrm{os}$ are the distances
between L and S, between O and L, and between O and S, respectively.
We assume $b_\mathrm{m}\ll D_\mathrm{ol}$ and $b_\mathrm{m}\ll D_\mathrm{ls}$ 
and we neglect $\theta=b/D_\mathrm{ol}$ and $\bar{\theta}=b/D_\mathrm{ls}$ in the lens equation.

\subsection{Light rays slightly outside of the photon sphere}
In the strong deflection limit $b \rightarrow b_\mathrm{m}+0$, 
the exact analytical expressions of $\bar{a}$ and $\bar{b}$ in the deflection angle (\ref{eq:al1}) in a Reissner-Nordstr\"{o}m spacetime
have been obtained as
\begin{equation}\label{eq:abar1}
\bar{a}=\frac{r_\mathrm{m}}{\sqrt{3mr_\mathrm{m}-4q^{2}}}
\end{equation}
and
\begin{eqnarray}\label{eq:bbar}
\bar{b}
&=&\bar{a} \log \left[ \frac{8(3mr_\mathrm{m}-4q^{2})^{3}}{m^{2}r_\mathrm{m}^{2}(mr_\mathrm{m}-q^{2})^{2}} \right. \nonumber\\
&&\left. \times \left(2\sqrt{mr_\mathrm{m}-q^{2}}-\sqrt{3mr_\mathrm{m}-4q^{2}} \right)^{2} \right] -\pi, \nonumber\\
\end{eqnarray}
respectively~\cite{Tsukamoto:2016jzh,Tsukamoto:2016oca}.
Here, $m$ and $q$ are mass and charge of the lensing object and $r_\mathrm{m}$ is the radius of the photon sphere given by 
\begin{equation}\label{eq:rm00}
r_\mathrm{m}=\frac{3m+\sqrt{9m^{2}-8q^{2}}}{2}
\end{equation}
for $0\leq q<3m/(2\sqrt{2})$. 
The analytic formulas recover the numerical results by Eiroa \textit{et al.}~\cite{Eiroa:2002mk} and Bozza~\cite{Bozza:2002zj} as shown in Refs~\cite{Tsukamoto:2016jzh,Tsukamoto:2016oca}.

From the deflection angle $\alpha(b)$~(\ref{eq:al1}) and $b=\theta D_\mathrm{ol}$, 
we obtain the positive solution of the lens equation~(\ref{eq:Lens1}) as $\theta=\theta_{n\mathrm{out}}(\beta)$, where 
\begin{equation}\label{eq:theta+10}
\theta_{n\mathrm{out}}(\beta)\equiv \theta_\mathrm{m} \left\{ 1+\exp\left[ \frac{\bar{b}-(1+2n)\pi+\beta}{\bar{a}} \right] \right\},
\end{equation}
and where $\theta_\mathrm{m}\equiv b_\mathrm{m}/D_\mathrm{ol}$ is the image angle of the photon sphere.
The magnification $\mu_{n\mathrm{out}}$ of the image is given by 
\begin{equation}\label{eq:magnification10}
\mu_{n\mathrm{out}}(\beta)=-\frac{D_\mathrm{os}^{2}}{D_\mathrm{ls}^{2}} s(\beta)\theta_{n\mathrm{out}}\frac{d\theta_{n\mathrm{out}}}{d\beta},
\end{equation}
where a function $s(\beta)$ for a point source is given by
\begin{equation}
s(\beta)=\frac{1}{\beta}.
\end{equation}

The function $s(\beta)$ for an uniform-luminous disk with a finite size on the observer's sky~\cite{Witt:1994,Nemiroff:1994uz,Alcock:1997fi} 
becomes an integral over the disk on a source plane:
\begin{equation}
s(\beta)
= \frac{1}{\pi \beta_\mathrm{s}^{2}} \int_{\mathrm{disk}} d\beta' d\phi,
\end{equation}
where $\beta'$ is a reduced radial coordinate which is divided by $D_\mathrm{ls}$ on the source plane, 
$\beta_\mathrm{s}\equiv R_\mathrm{s}/D_\mathrm{ls}$ is the reduced radius of the Sun, where $R_\mathrm{s}$ is the radius of the Sun,
and where $\phi$ is an azimuthal coordinate around the origin on the source plane.
By fixing the origin of the coordinates on the source plane, which is the intersection point of an axis $\beta=0$ and the source plane,
$s(\beta)$ can be rewritten as
\begin{eqnarray}
s(\beta)
&=& \frac{2}{\pi \beta_\mathrm{s}^{2}} \left[ \pi(\beta_\mathrm{s}-\beta) \right. \nonumber\\
&&\left.+\int^{\beta+\beta_\mathrm{s}}_{-\beta+\beta_\mathrm{s}} \arccos  \frac{\beta^{2}+\beta'^{2}-\beta^{2}_\mathrm{s}}{2\beta \beta'} d\beta' \right]
\end{eqnarray}
for $\beta \leq \beta_\mathrm{s}$ and
\begin{eqnarray}
s(\beta)
=\frac{2}{\pi \beta_\mathrm{s}^{2}} \int^{\beta+\beta_\mathrm{s}}_{\beta-\beta_\mathrm{s}} \arccos \frac{\beta^{2}+\beta'^{2}-\beta^{2}_\mathrm{s}}{2\beta \beta'} d\beta' 
\end{eqnarray}
for $\beta_\mathrm{s} \leq \beta$.
If the photon sphere, the observer, and the Sun are perfectly aligned,
$s(\beta)$ is given by
\begin{equation}
s(0)=\frac{2}{\beta_\mathrm{s}}.
\end{equation}
From Eqs.~(\ref{eq:theta+10}) and (\ref{eq:magnification10}), 
its magnification $\mu_{n\mathrm{out}}(\beta)$ is obtained as
\begin{eqnarray}
\mu_{n\mathrm{out}}(\beta)
&=&-\frac{D_\mathrm{os}^{2}}{D_\mathrm{ls}^{2}}\frac{\theta_\mathrm{m}^{2}e^{\left[\bar{b}-(1+2n)\pi\right]/\bar{a}}}{\bar{a}} \nonumber\\
&&\times \left\{ 1+e^{\left[\bar{b}-(1+2n)\pi\right]/\bar{a}} \right\} s(\beta).
\end{eqnarray}

Notice that the lens equation has a negative solution
$\theta \sim -\theta_{n\mathrm{out}}(\beta)$
and its magnification is given by $-\mu_{n\mathrm{out}}(\beta)$ approximately.
The total magnification $\mu_\mathrm{totout}(\beta)$ of the couples of the images from $n=0$ to $\infty$ is given by
\begin{eqnarray}
\mu_\mathrm{totout}(\beta)
&\equiv& 2 \sum_{n=0}^{\infty} \left| \mu_{n\mathrm{out}}(\beta) \right|  \nonumber\\
&=& 2\frac{D_\mathrm{os}^{2}}{D_\mathrm{ls}^{2}}
\frac{\theta_\mathrm{m}^{2}}{\bar{a}}\left|s(\beta)\right| 
\left[ \frac{e^{(\bar{b}-\pi)/\bar{a}}}{1-e^{-2\pi/\bar{a}}}+\frac{e^{2(\bar{b}-\pi)/\bar{a}}}{1-e^{-4\pi/\bar{a}}} \right] \nonumber\\
\end{eqnarray}
and it gives, in the perfectly aligned case,
\begin{equation}\label{eq:aligned_magnification}
\mu_\mathrm{totout}(0)
=4\frac{D_\mathrm{os}^{2}}{D_\mathrm{ls}^{2}}\frac{\theta_\mathrm{m}^{2}}{\bar{a}\beta_\mathrm{s}}
\left[ \frac{e^{(\bar{b}-\pi)/\bar{a}}}{1-e^{-2\pi/\bar{a}}}+\frac{e^{2(\bar{b}-\pi)/\bar{a}}}{1-e^{-4\pi/\bar{a}}} \right].
\end{equation}

\begin{figure}[htbp]
\begin{center}
\includegraphics[width=80mm]{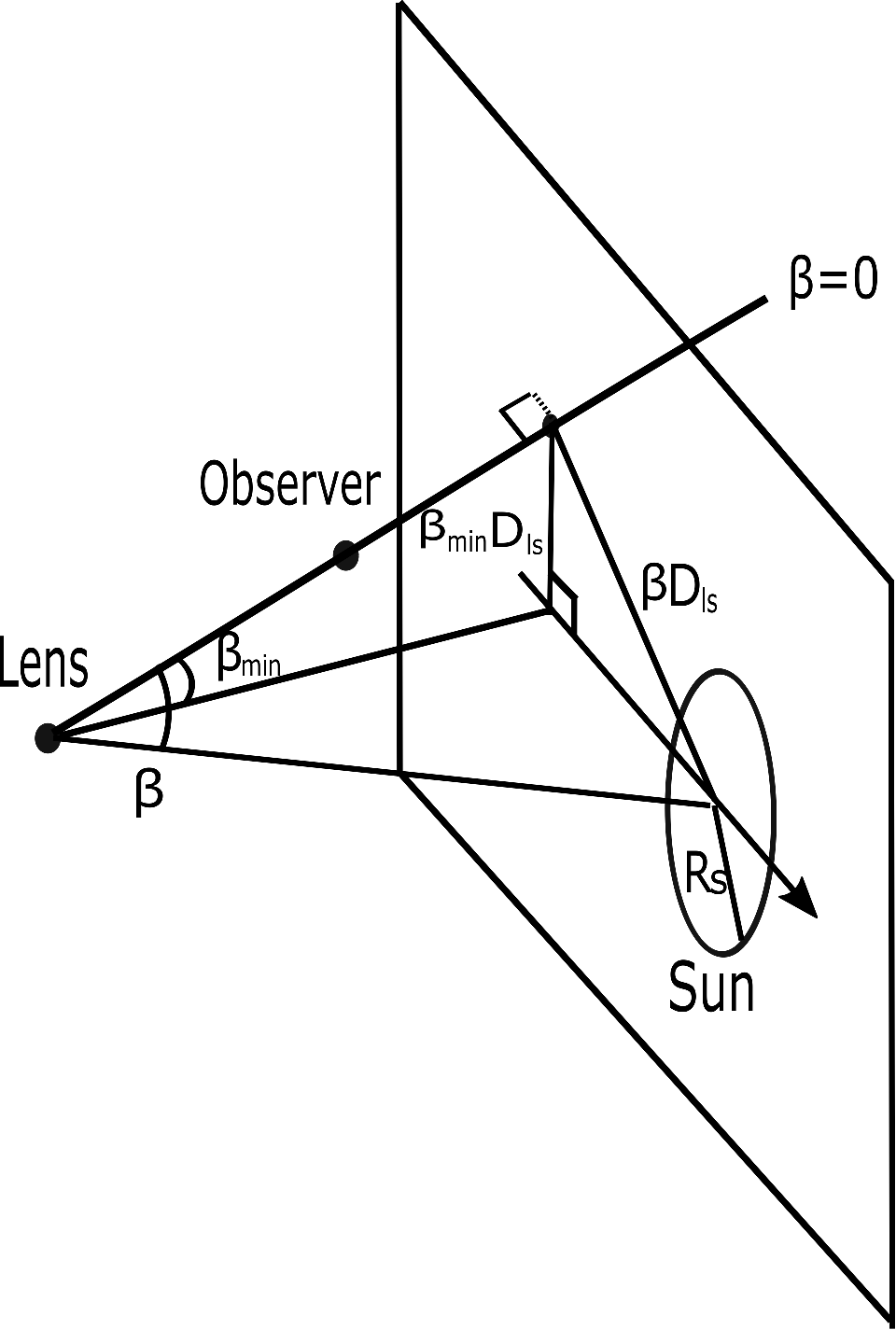}
\end{center}
\caption{Retrolensing. 
The Sun is on the source plane and the plane is orthogonal to the optical axis $\beta=0$. 
$\beta_\mathrm{min}$ denotes the smallest source angle. 
We assume that the Sun moves with the orbital velocity $v=30$km/s on the source plane.}
\label{microlens2}
\end{figure}

We consider the retrolensing of the sunlight reflected by the photon sphere at $D_\mathrm{ol}=0.01$pc away
as shown in Fig.~\ref{microlens2}.
We assume that the Sun moves with the orbital velocity $v=30$km/s on a source plane
and with the closest separation $\beta_\mathrm{min}$ between the center of the Sun disk 
and the axis $\beta=0$ on the source plane as shown Fig.~\ref{microlens2}.

\begin{figure}[htbp]
\begin{center}
\includegraphics[width=85mm]{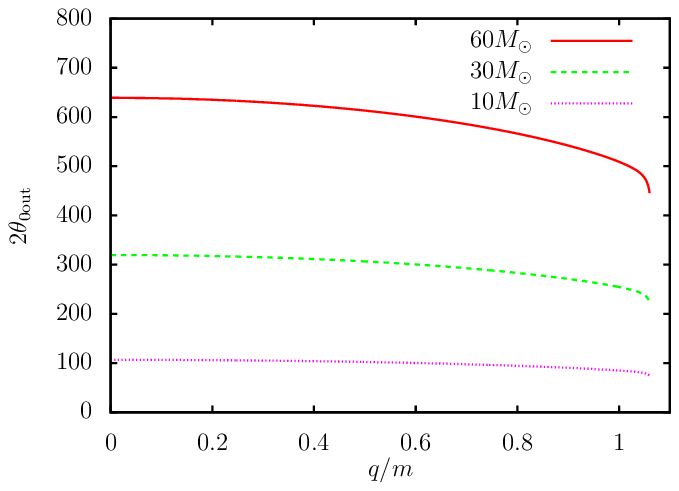}
\end{center}
\caption{The image separation $2\theta_{0\mathrm{out}}$ of outer images as a function of $q/m$.
Solid (red), dashed (green), and dotted (magenta) curves denote the image separation in the cases of $m=60M_{\odot}$, $30M_{\odot}$, and $10M_{\odot}$, 
respectively.
The distance to the photon sphere is $D_\mathrm{ol}=0.01$pc and the source angle is $\beta=0$ and the winding number is $n=0$.
}
\label{imageout}
\end{figure}

The separation between the image $\theta_{n\mathrm{out}}$ and its paired image is given by
\begin{equation}
2\theta_{n\mathrm{out}}
=2\theta_\mathrm{m} \left\{ 1+\exp\left[ \frac{\bar{b}-(1+2n)\pi+\beta}{\bar{a}} \right] \right\}
\end{equation}
and the separation of the image angles with the winding number $n=0$ as a function of $q/m$ is plotted in Fig.~\ref{imageout}. 

\subsection{Light rays slightly inside of the photon sphere}
In the case of light rays reflected by a potential barrier near the antiphoton sphere for $m<q<3m/(2\sqrt{2})$,
the exact forms of $\bar{c}$ and $\bar{d}$ of the deflection angle~(\ref{eq:al2}) in the strong deflection limit $b \rightarrow b_\mathrm{m}-0$ 
are obtained by Tsukamoto~\cite{Tsukamoto:2021fsz} as 
\begin{equation}
\bar{c}\equiv \frac{2r_\mathrm{m}}{\sqrt{3mr_\mathrm{m}-4q^2}}
\end{equation}
and
\begin{eqnarray}
\bar{d}
&=&\bar{c} \log \left[ \frac{16\left(3 m r_\mathrm{m}-4 q^2\right)^3 (r_\mathrm{m}-r_\mathrm{c})}{2(mr_\mathrm{m}-q^2)+\sqrt{(3mr_\mathrm{m}- 4q^2)(mr_\mathrm{m}-q^2)}         } \right. \nonumber\\
&&\times \left. \frac{1}{(mr_\mathrm{m}-q^2) \left\{ mr_\mathrm{m}(r_\mathrm{m}+2r_\mathrm{c})-2q^2(r_\mathrm{m}+r_\mathrm{c}) \right\} } \right] -\pi, \nonumber\\
\end{eqnarray}
where $r_\mathrm{c}$ is the smaller positive zero point of the effective potential 
of the light ray with the impact critical parameter $b_\mathrm{m}$
given by
\begin{equation}\label{eq:rc00}
r_\mathrm{c}=\frac{r_\mathrm{m}\left( \sqrt{mr_\mathrm{m}}-\sqrt{mr_\mathrm{m}-q^2} \right)}{\sqrt{mr_\mathrm{m}-q^2}}
\end{equation}
for $m<q<3m/(2\sqrt{2})$.

For the deflection angle (\ref{eq:al2}), the positive solution of the lens equation~(\ref{eq:Lens1}) is obtained by
$\theta=\theta_{n\mathrm{in}}(\beta)$, where $\theta_{n\mathrm{in}}(\beta)$ is given by
\begin{equation}\label{eq:theta01}
\theta_{n\mathrm{in}}(\beta)\equiv 
\frac{\theta_{\mathrm{m}}}{\sqrt{1+\exp \left[ \frac{\bar{d}-(1+2n)\pi+\beta}{\bar{c}} \right] }}
\end{equation}
and its magnification is given by
\begin{equation}
\mu_{n\mathrm{in}}(\beta)
=\frac{D_\mathrm{os}^{2}}{2D_\mathrm{ls}^{2}}
\frac{\theta_\mathrm{m}^{2}\exp\left[ \frac{\bar{d}-(1+2n)\pi}{\bar{c}} \right]}{\bar{c}\left\{ 1+\exp\left[ \frac{\bar{d}-(1+2n)\pi}{\bar{c}} \right] \right\}^2}s(\beta).
\end{equation}
A negative solution of the lens equation is $\theta=\theta_{-n\mathrm{in}}(\beta)$, where 
\begin{equation}\label{eq:negative_image}
\theta_{-n\mathrm{in}}(\beta)= -\theta_{n\mathrm{in}}(-\beta) \sim -\theta_{n\mathrm{in}}(\beta)
\end{equation}
and its magnification is given by $-\mu_{n\mathrm{in}}(\beta)$.
The total magnification $\mu_\mathrm{totin}(\beta)$ of the pair images from $n=0$ to $\infty$ is given by
\begin{eqnarray}
\mu_\mathrm{totin}(\beta)
&=& 2 \sum_{n=0}^{\infty} \left|  \mu_{n\mathrm{in}}(\beta) \right| \nonumber\\
&=& \sum_{n=0}^{\infty} 
\frac{D_\mathrm{os}^{2}}{D_\mathrm{ls}^{2}}
\frac{\theta_\mathrm{m}^{2}\exp\left[ \frac{\bar{d}-(1+2n)\pi}{\bar{c}} \right]}{\bar{c}\left\{ 1+\exp\left[ \frac{\bar{d}-(1+2n)\pi}{\bar{c}} \right] \right\}^2}
\left|s(\beta)\right| \nonumber\\
\end{eqnarray}
and it gives, in the perfectly aligned case,
\begin{equation}\label{eq:aligned_magnification2}
\mu_\mathrm{totin}(0)
=2\sum_{n=0}^{\infty} \frac{D_\mathrm{os}^{2}}{D_\mathrm{ls}^{2}}
\frac{\theta_\mathrm{m}^{2}\exp\left[ \frac{\bar{d}-(1+2n)\pi}{\bar{c}} \right]}{\bar{c}\left\{ 1+\exp\left[ \frac{\bar{d}-(1+2n)\pi}{\bar{c}} \right] \right\}^2\beta_\mathrm{s}}.
\end{equation}

\begin{figure}[htbp]
\begin{center}
\includegraphics[width=85mm]{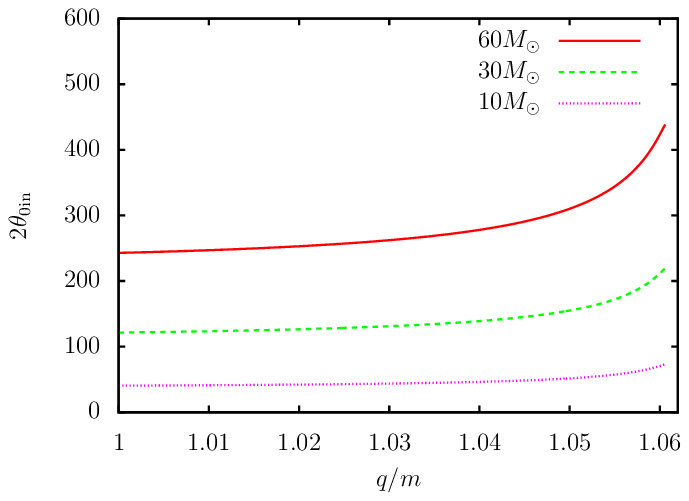}
\end{center}
\caption{The image separation  $2\theta_{0\mathrm{in}}$ of the inner images as a function of $q/m$.
The curve types, $m$, $D_\mathrm{ol}$, $\beta$, and $n$ are the same as the ones of Fig.~\ref{imageout}.
}
\label{imagein}
\end{figure}

The image separation between $\theta_{n\mathrm{in}}$ and its pair is given by
\begin{equation}
2\theta_{n\mathrm{in}}
=\frac{2\theta_{\mathrm{m}}}{\sqrt{1+\exp \left[ \frac{\bar{d}-(1+2n)\pi+\beta}{\bar{c}} \right] }}
\end{equation}
and the case of $n=0$ is shown in Fig.~\ref{imagein}. 

\subsection{Percent errors of the deflection angle in the strong deflection limits}
The deflection angle $\alpha$ of the ray is obtained as    
\begin{equation}\label{eq:deflection00}
\alpha=2 \int^\infty_{r_0} \frac{dr}{r \sqrt{ \frac{r^2}{b^2} -A }}-\pi,
\end{equation}
where $r_0$ is the radial coordinate of the reflection point of the light ray, 
$b=b(r_0)$ is the impact parameter of the ray and it can be expressed by 
\begin{equation}
b=\pm \sqrt{\frac{r_0^2}{A(r_0)}}
\end{equation}
and 
$A(r)$ is defined by 
\begin{equation}
A(r)=1-\frac{2m}{r}+\frac{q^2}{r^2}.
\end{equation}
We consider the positive impact parameter only unless we comment on the negative impact parameter.
See Appendix A for a short review on the Reissner-Nordstr\"{o}m spacetime and the deflection angle.

The percent errors of deflection angle calculated by 
\begin{eqnarray}\label{eq:per11}
\frac{\alpha \mathrm{\:  of \: Eq. \: (1.1)}-\alpha \mathrm{\: of \: Eq. \: (2.30)}}{\alpha \mathrm{ \:of \: Eq. \: (2.30)}} \times 100 
\end{eqnarray}
and
\begin{eqnarray}\label{eq:per12}
\frac{\alpha \mathrm{\:  of \: Eq. \: (1.2)}-\alpha \mathrm{\: of \: Eq. \: (2.30)}}{\alpha \mathrm{ \:of \: Eq. \: (2.30)}} \times 100 
\end{eqnarray}
as a function of $\alpha$ of Eq.~(\ref{eq:deflection00}) 
are shown in Figs. 5 and 6, respectively.
\begin{figure}[htbp]
\begin{center}
\includegraphics[width=80mm]{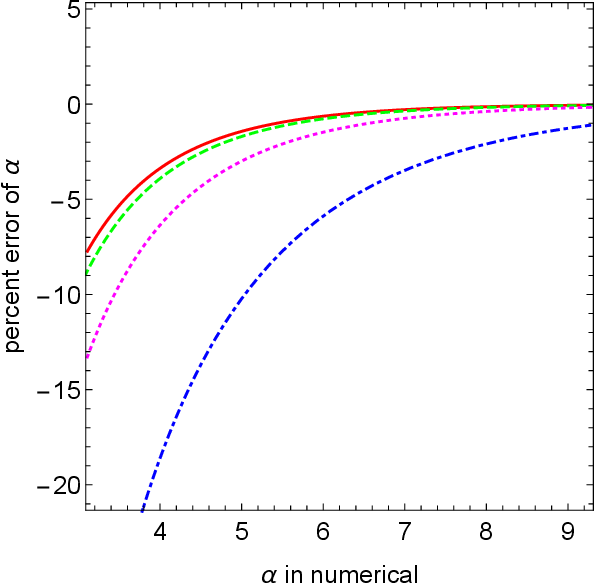}
\end{center}
\caption{The percent error (\ref{eq:per11}) of $\alpha$ of Eq. (\ref{eq:al1}) as a function of $\alpha$ of Eq.(\ref{eq:deflection00}) calculated in numerical for outer images.
A solid (red), dashed (green), dotted (magenta), and dot-dashed (blue) curves denote the percent errors
for $q/m=1.001$, $1.01$, $1.03$, and $1.05$, respectively.}
\label{fig:5}
\end{figure}
\begin{figure}[htbp]
\begin{center}
\includegraphics[width=80mm]{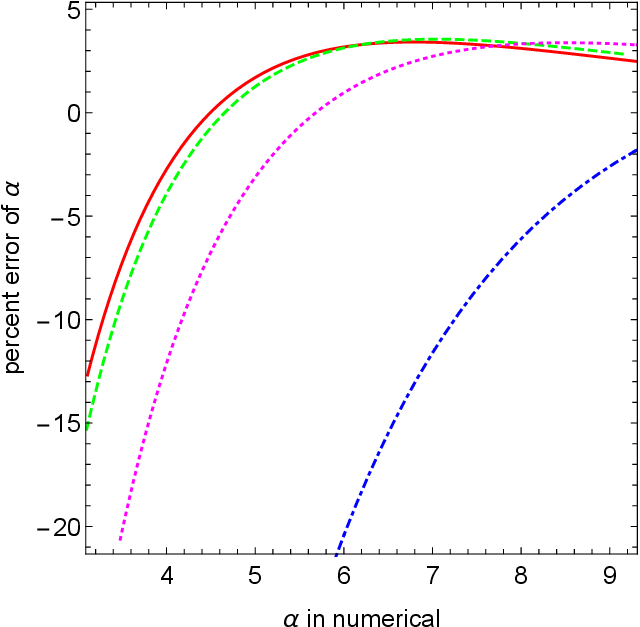}
\end{center}
\caption{The percent error (\ref{eq:per12}) of $\alpha$ of Eq. (\ref{eq:al2}) as a function of $\alpha$ of Eq.(\ref{eq:deflection00}) calculated numerically for inner images.
The types of curves for the given $q/m$ are the same as Fig.~\ref{fig:5}.}
\label{fig:6}
\end{figure}
Note that the percent errors (\ref{eq:per11}) and (\ref{eq:per12}) of the deflection angle $\alpha\sim \pi$ for $q/m\sim 0.1$ 
are about $-10$ percent and that the ones of $\alpha\sim 3\pi$ become a few percent. 
In the almost marginally unstable photon sphere case with $q/m \lesssim 3/(2\sqrt{2}),$
the absolute values of the percent errors of the deflection angle $\alpha \sim \pi$ are huge.
This implies that the deflection angles diverge nonlogarithmically in the marginally unstable photon sphere case with $q/m=3/(2\sqrt{2})$.

\subsection{Retrolensing light curves}
\begin{figure}[htbp]
\begin{center}
\includegraphics[width=85mm]{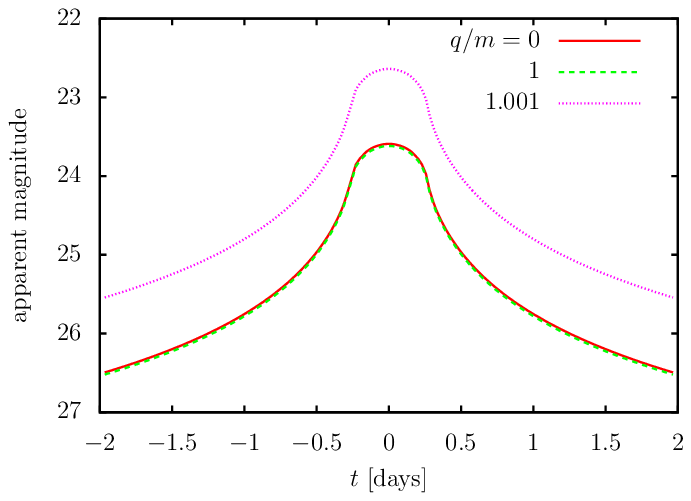}
\end{center}
\caption{Light curves with the charge $q/m=0$, $1$, and $1.001$ are denoted by (red) solid, (green) dashed, and (magenta) dotted curves, respectively. 
We set the distance to the lens $D_\mathrm{ol}=0.01$pc, the mass $m=30M_{\odot}$, and the smallest source angle $\beta_\mathrm{min}=0$.}
\label{lc1}
\end{figure}

\begin{figure}[htbp]
\begin{center}
\includegraphics[width=85mm]{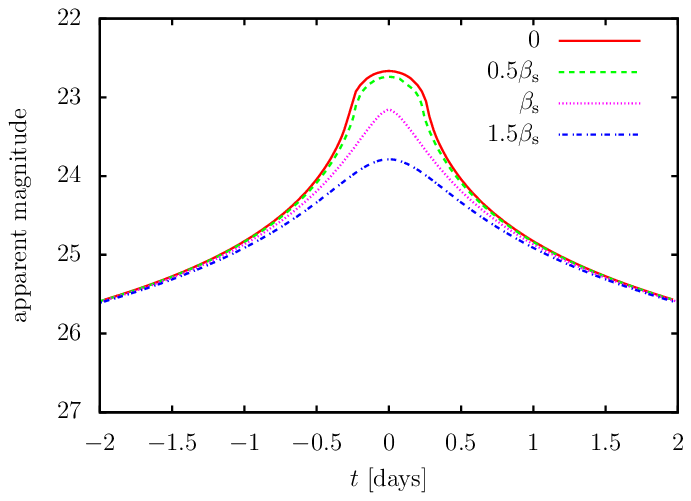}
\end{center}
\caption{Light curves with the smallest source angle $\beta_\mathrm{min}=0$, $0.5\beta_\mathrm{s}$, $\beta_\mathrm{s}$, and $1.5\beta_\mathrm{s}$
 are denoted by (red) solid, (green) dashed, (magenta) dotted, and (blue) dash-dotted curves, respectively.
We set the distance to the lens $D_\mathrm{ol}=0.01$pc, the mass $m=30M_{\odot}$, 
and the charge $q/m=1.01$.}
\label{lc2}
\end{figure}

\begin{figure}[htbp]
\begin{center}
\includegraphics[width=85mm]{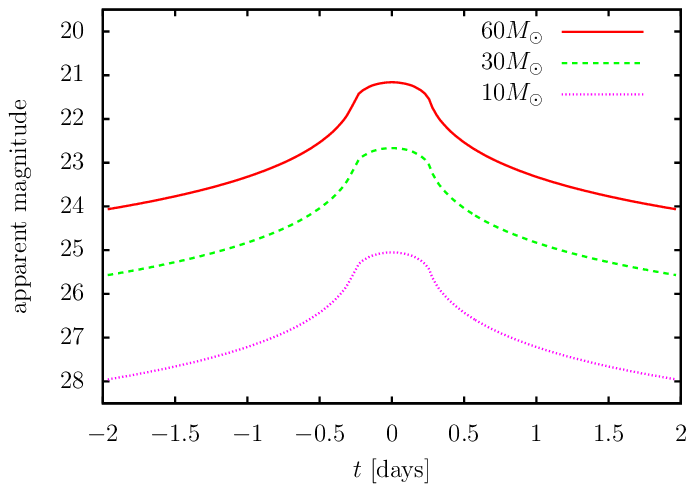}
\end{center}
\caption{Light curves with the mass $m=60M_{\odot}$, $30M_{\odot}$, and $10M_{\odot}$ 
 are denoted by (red) solid, (green) dashed, and (magenta) dotted curves, respectively.
We set the distance to the lens $D_\mathrm{ol}=0.01$pc, the charge $q/m=1.01$, and the smallest source angle $\beta_\mathrm{min}=0$.}
\label{lc3}
\end{figure}

\begin{figure}[htbp]
\begin{center}
\includegraphics[width=85mm]{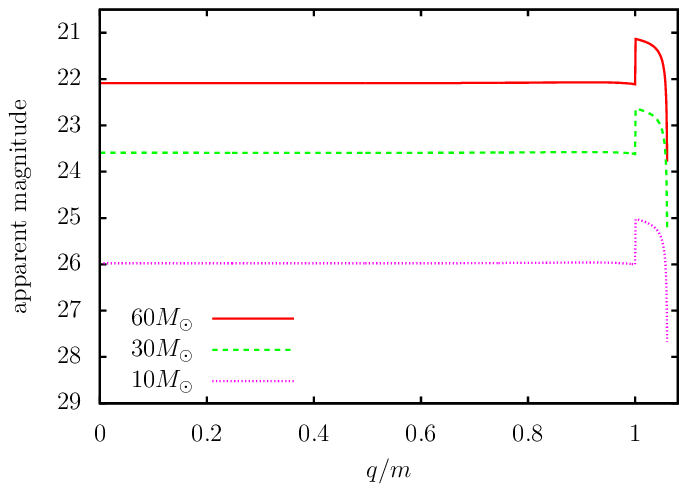}
\end{center}
\caption{The apparent magnitudes of the peak
of a photon sphere at a distance $D_\mathrm{ol}=0.01$pc as a function of $q/m$ in the perfectly aligned case ($\beta_\mathrm{min}=0$). 
The apparent magnitudes of the peak 
for the mass with $m=60M_{\odot}$, $30M_{\odot}$, and $10M_{\odot}$ are shown as 
 (red) solid, (green) dashed, and (magenta) dotted curves, respectively.
Note that the error of the strong deflection limit analysis at almost marginally unstable case $q/m\sim 1.06$ is huge. }
\label{pk}
\end{figure}

The light curves of retrolensing by the photon sphere in the Reissner-Nordstr\"{o}m black hole and naked singularity spacetimes are shown 
in Figs.~\ref{lc1}-\ref{lc3}.
Figure~\ref{pk} shows the the apparent magnitude of the light curves at the peak in the perfectly aligned case as a function of $q/m$. 
Notice that light curves consist of only light rays reflected slightly outside of the photon sphere around the black hole for $q/m\leq 1$ while 
the ones consist of both light rays reflected slightly outside and inside of the photon sphere around the naked singularity for $1<q/m<3/(2\sqrt{2})$.

\section{Discussion and Conclusion}
We have investigated the retrolensing of the sunlight reflected by the photon sphere and by the potential barrier near the antiphoton sphere 
in the Reissner-Nordstr\"{o}m black hole and naked singularity spacetimes 
by using the exact forms of the deflection angles in the strong deflection limits $b\rightarrow b_\mathrm{m}-0$ and $b\rightarrow b_\mathrm{m}+0$. 
Note that we can apply the formula for the retrolensing by the Reissner-Nordstr\"{o}m black hole investigated by Tsukamoto and Gong~\cite{Tsukamoto:2016oca}
to the retrolensing of the rays reflected at slightly outside of the photon sphere around the naked singularity for $1<q/m<3/(2\sqrt{2})$.
The retrolensing by the photon sphere around the Reissner-Nordstr\"{o}m naked singularity can be brighter 
than the one around the Reissner-Nordstr\"{o}m black hole 
because of the light rays reflected by the potential barrier near the antiphoton sphere.

The light curves of the retrolensing have characteristic shapes as shown in Figs.~\ref{lc1}-\ref{lc3} because of the spherical symmetry and the disk shape of the Sun.
The time-symmetrical shape of the light curves helps us to distinguish the retrolensing from other phenomena of light emissions,
and we would distinguish the events of retrolensing from the other events since 
the retrolensing light curves have precise solar spectra and they can be observed on the ecliptic~\cite{Holz:2002uf}.

We do not consider the case of $q/m=3/(2\sqrt{2})$ in this paper.
For $q/m=3/(2\sqrt{2})$, the effective potential of the light ray with the critical impact parameter $b=b_\mathrm{m}$ 
gives $V(r_\mathrm{m})=V^{\prime}(r_\mathrm{m})=V^{\prime \prime}(r_\mathrm{m})=0$ and $V^{\prime \prime \prime}(r_\mathrm{m})<0$,
where the prime is a differentiation with respect to the radial coordinate $r$,  
and the photon sphere and antiphoton photon sphere come together as one marginally unstable photon sphere at $r=r_\mathrm{m}=r_{\mathrm{aps}}=3m/2$. 
In this case, the deflection angle diverges nonlogarithmically and we cannot apply the formulas obtained in this paper to the marginally unstable photon sphere. 
The deflection angle in the strong deflection limit $b\rightarrow b_\mathrm{m}+0$ of the marginally unstable photon sphere of the Reissner-Nordstr\"{o}m spacetime
and its application to a usual lens configuration have been investigated by Tsukamoto~\cite{Tsukamoto:2020iez}.
The deflection angle of the marginally unstable photon sphere in the strong deflection limit $b\rightarrow b_\mathrm{m}-0$ 
and retrolensing by the marginally unstable photon sphere in the strong deflection limits
$b\rightarrow b_\mathrm{m}-0$ and $b\rightarrow b_\mathrm{m}+0$ are left as future work.

%
\appendix
\section{Deflection angle in the Reissner-Nordstr\"{o}m spacetime}
In this appendix, we briefly review the deflection angle of light rays in strong deflection limits in a Reissner-Nordstr\"{o}m spacetime
(see Refs.~\cite{Tsukamoto:2016jzh,Tsukamoto:2016oca,Tsukamoto:2021fsz} for details).
The Reissner-Nordstr\"{o}m spacetime has a line element 
\begin{equation}\label{eq:metric}
ds^2=-A(r)dt^2+\frac{dr^2}{A(r)}+r^2 (d\vartheta^2+\sin^2 \vartheta d\varphi^2),
\end{equation}
where $A(r)$ is given by
\begin{equation}
A(r)=1-\frac{2m}{r}+\frac{q^2}{r^2}
\end{equation}
and $m\geq 0$ is an Arnowitt-Deser-Misner mass and $q\geq 0$ is an electrical charge
and it has time-translational and axial Killing vectors $t^\mu \partial_\mu=\partial_t$ and $\varphi^\mu \partial_\mu=\partial_\varphi$ 
because of its stationarity and axisymmetry, respectively.
There is an event horizon $r=r_\mathrm{H}$, where 
\begin{equation}
r_\mathrm{H}\equiv m+\sqrt{m^{2}-q^{2}},
\end{equation}
for $q\leq m$ while the spacetime has a naked singularity for $m<q$.
Without loss of generality, we can assume $\vartheta=\pi/2$ because of spherical symmetry.

By using a wave vector $k^\mu\equiv \dot{x}^\mu$, where the dot denotes a differentiation with respect to an affine parameter, 
the trajectory of a light ray is expressed by $k^\mu k_\mu=0$ and it can be rewritten as
\begin{equation}\label{eq:trajectory}
-A\dot{t}^2+\frac{\dot{r}^2}{A}+r^2 \dot{\varphi}^2=0.
\end{equation}
The light ray can be characterized by its closest distant or its reflectional point $r=r_0$ 
and an equation
\begin{equation}\label{eq:trajectory0}
A_0\dot{t}^2_0=r^2_0 \dot{\varphi}^2_0
\end{equation}
holds at the reflectional point.
Here and hereinafter functions with subscript $0$ denote the functions at the reflectional point $r=r_0$.
The impact parameter of a light is given by
\begin{equation}
b(r_0)\equiv \frac{L}{E}=\frac{r_0^2 \dot{\varphi}_0}{A_0\dot{t}_0},
\end{equation}
where $E\equiv -g_{\mu \nu} t^\mu k^\nu=A\dot{t}$ and 
$L \equiv g_{\mu \nu} \varphi^\mu k^\nu=r^2 \dot{\varphi}$ are conserved energy and conserved angular momentum of the ray, respectively.
From Eq.~(\ref{eq:trajectory0}),
the impact parameter can be expressed by 
\begin{equation}
b=\pm \sqrt{\frac{r_0^2}{A_0}}.
\end{equation}
From Eq.~(\ref{eq:trajectory}), the radial motion of the light is given by 
\begin{equation}
\dot{r}^2+\frac{V(r)}{E^2}=0,
\end{equation}
where $V(r)$ is an effective potential defined by
\begin{equation}
V(r)\equiv \frac{Ab^2}{r^2}-1.
\end{equation}
The ray can exist in a region $V(r)\leq 0$.
There is a photon sphere at $r=r_\mathrm{m}$, where $r_\mathrm{m}$ is given by
\begin{equation}\label{eq:rm1}
r_\mathrm{m}=\frac{3m+\sqrt{9m^{2}-8q^{2}}}{2}
\end{equation}
for $0\leq q<3m/(2\sqrt{2})$ 
and an antiphoton sphere at $r=r_{\mathrm{aps}}$, where $r_{\mathrm{aps}}$ is
\begin{equation}\label{eq:raps}
r_\mathrm{aps}=\frac{3m-\sqrt{9m^{2}-8q^{2}}}{2}
\end{equation}
for $m<q<3m/(2\sqrt{2})$. 
Note that $V(r_\mathrm{m})=V^{\prime}(r_\mathrm{m})=0$ and $V^{\prime \prime}(r_\mathrm{m})<0$ hold for the critical impact parameter $b=b_\mathrm{m} \equiv b(r_0=r_\mathrm{m})$ and that
$V(r_\mathrm{aps})=V^{\prime}(r_\mathrm{aps})=0$ and $V^{\prime \prime}(r_\mathrm{aps})>0$ hold for the critical impact parameter.
Notice that the prime is the differentiation with respect to the radial coordinate $r$.
We define the smaller positive zero point $r=r_c$ of the effective potential of the light ray with the impact critical parameter $b_\mathrm{m}$ for $m<q<3m/(2\sqrt{2})$.
We obtain $r_\mathrm{c}$ as
\begin{equation}\label{eq:rc}
r_\mathrm{c}=\frac{r_\mathrm{m}\left( \sqrt{mr_\mathrm{m}}-\sqrt{mr_\mathrm{m}-q^2} \right)}{\sqrt{mr_\mathrm{m}-q^2}}.
\end{equation}
We note that $b(r_0=r_\mathrm{c})=b_\mathrm{m}$.
A plot of the radial coordinates of the photon sphere $r_\mathrm{m}$, 
an antiphoton sphere $r_\mathrm{aps}$, 
an event horizon $r_\mathrm{H}$, 
and the smaller positive zero point $r_\mathrm{c}$ of effective potential is shown in Ref.~\cite{Tsukamoto:2021fsz}. 

We consider that a light ray comes from at a spatial infinity and that it is reflected at $r=r_0$. 
From Eq.~(\ref{eq:trajectory}), the deflection angle $\alpha$ of the ray is obtained as    
\begin{equation}\label{eq:deflection}
\alpha=2 \int^\infty_{r_0} \frac{dr}{r \sqrt{ \frac{r^2}{b^2} -A }}-\pi.
\end{equation}


\begin{thebibliography}{99}

\bibitem{Schneider_Ehlers_Falco_1992}
P. Schneider, J. Ehlers, and E. E. Falco,
\textit{Gravitational Lenses} (Springer-Verlag, Berlin, 1992).

\bibitem{Hagihara_1931} 
Y.~Hagihara, 
Jpn.\ J.\ Astron.\ Geophys., {\bf 8}, 67 (1931).

\bibitem{Darwin_1959}
C. Darwin,
Proc. R. Soc. Lond. A {\bf 249}, 180 (1959).

\bibitem{Atkinson_1965}
R.~d'~E. Atkinson,
Astron. J. {\bf 70}, 517 (1965).

\bibitem{Luminet_1979}
J.-P. Luminet,  
Astron. Astrophys. {\bf 75}, 228 (1979).

\bibitem{Ohanian_1987}
H. C. Ohanian, 
Am. J. Phys. {\bf 55}, 428 (1987).

\bibitem{Nemiroff_1993}
R. J. Nemiroff,  
Am. J. Phys. {\bf 61}, 619 (1993).

\bibitem{Virbhadra:1998dy}
K.~S.~Virbhadra, D.~Narasimha, and S.~M.~Chitre,
Astron. Astrophys. {\bf 337}, 1 (1998).

\bibitem{Frittelli_Kling_Newman_2000}
S. Frittelli, T. P. Kling, and E. T. Newman,
Phys. Rev. D {\bf 61}, 064021 (2000).

\bibitem{Virbhadra_Ellis_2000}
K. S. Virbhadra and G. F. R. Ellis,
Phys. Rev. D {\bf 62}, 084003 (2000).

\bibitem{Bozza_Capozziello_Iovane_Scarpetta_2001}
V. Bozza, S. Capozziello, G. Iovane, and G. Scarpetta,
Gen. Relativ. Gravit. {\bf 33}, 1535 (2001).

\bibitem{Bozza:2002zj} 
  V.~Bozza,
  Phys.\ Rev.\ D {\bf 66}, 103001 (2002).

\bibitem{Virbhadra:2002ju}
K.~S.~Virbhadra and G.~F.~R.~Ellis,
Phys. Rev. D {\bf 65}, 103004 (2002).

\bibitem{Perlick:2003vg}
V.~Perlick,
Phys. Rev. D {\bf 69}, 064017 (2004).

\bibitem{Virbhadra:2008ws}
K.~S.~Virbhadra,
Phys. Rev. D {\bf 79}, 083004 (2009).

\bibitem{Bozza_2010}
V. Bozza,
Gen. Relativ. Gravit. {\bf 42}, 2269 (2010).

\bibitem{Tsukamoto:2016zdu}
  N.~Tsukamoto and T.~Harada,
  Phys.\ Rev.\ D {\bf 95}, 024030 (2017).

\bibitem{Shaikh:2019jfr} 
  R.~Shaikh, P.~Banerjee, S.~Paul, and T.~Sarkar,
  JCAP {\bf 1907}, 028 (2019).

\bibitem{Shaikh:2019itn} 
  R.~Shaikh, P.~Banerjee, S.~Paul, and T.~Sarkar,
  Phys.\ Rev.\ D {\bf 99}, 104040 (2019).

\bibitem{Tsukamoto:2020uay}
N.~Tsukamoto,
Phys. Rev. D {\bf 101}, 104021 (2020).

\bibitem{Tsukamoto:2020iez}
N.~Tsukamoto,
Phys. Rev. D {\bf 102}, 104029 (2020).

\bibitem{Paul:2020ufc}
S.~Paul,
Phys. Rev. D {\bf 102}, 064045 (2020).

\bibitem{Perlick_2004_Living_Rev}
V. Perlick,
Living Rev. Relativity {\bf7}, 9 (2004).

\bibitem{Claudel:2000yi} 
  C.~M.~Claudel, K.~S.~Virbhadra, and  G.~F.~R.~Ellis,
  J.\ Math.\ Phys.\  {\bf 42}, 818 (2001).

\bibitem{Perlick:2021aok}
V.~Perlick and O.~Y.~Tsupko,
[arXiv:2105.07101 [gr-qc]].

\bibitem{Abbott:2016blz}
  B.~P.~Abbott {\it et al.} [LIGO Scientific and Virgo Collaborations],
  Phys.\ Rev.\ Lett.\  {\bf 116}, 061102 (2016).
  
\bibitem{Akiyama:2019cqa} 
  K.~Akiyama {\it et al.} [Event Horizon Telescope Collaboration],
  Astrophys.\ J.\  {\bf 875}, L1 (2019).

\bibitem{Keir:2014oka} 
  J.~Keir,
  Class.\ Quant.\ Grav.\  {\bf 33}, 135009 (2016).

\bibitem{Cardoso:2014sna} 
  V.~Cardoso, L.~C.~B.~Crispino, C.~F.~B.~Macedo, H.~Okawa, and P.~Pani,
  Phys.\ Rev.\ D {\bf 90}, 044069 (2014).

\bibitem{Cunha:2017qtt} 
  P.~V.~P.~Cunha, E.~Berti, and C.~A.~R.~Herdeiro,
  Phys.\ Rev.\ Lett.\  {\bf 119}, 251102 (2017).

\bibitem{Hod:2017xkz} 
  S.~Hod,
  Phys.\ Lett.\ B {\bf 727}, 345 (2013);

\bibitem{Sanchez:1977si} 
  N.~G.~Sanchez,
  Phys.\ Rev.\ D {\bf 18}, 1030 (1978);

  \bibitem{Press:1971wr} 
  W.~H.~Press,
  Astrophys.\ J.\  {\bf 170}, L105 (1971);
  I.~Z.~Stefanov, S.~S.~Yazadjiev, and G.~G.~Gyulchev,
  Phys.\ Rev.\ Lett.\  {\bf 104}, 251103 (2010);

\bibitem{Abramowicz:1990cb} 
  M.~A.~Abramowicz,
Mon.\ Not.\ Roy.\ Astr.\ Soc.\ {\bf 245}, 733 (1990);
W. Hasse and V. Perlick,
Gen. Relativ. Gravit. {\bf 34}, 415 (2002).

\bibitem{Koga:2016jjq} 
  Y.~Koga and T.~Harada,
  Phys.\ Rev.\ D {\bf 94}, 044053 (2016);

\bibitem{Barcelo:2000ta}
C.~Barcelo and M.~Visser,
Nucl. Phys. B {\bf 584}, 415 (2000);
Y.~Koga,
Phys. Rev. D {\bf 101} 104022 (2020).

\bibitem{Ames_1968} 
W. L. Ames and K. S. Thorne, 
Astrophys.\ J. {\bf 151}, 659 (1968).

\bibitem{Gibbons:2016isj}
G.~W.~Gibbons and C.~M.~Warnick,
Phys. Lett. B {\bf 763}, 169 (2016);
  P.~V.~P.~Cunha, C.~A.~R.~Herdeiro, and E.~Radu,
  Phys.\ Rev.\ D {\bf 96}, 024039 (2017);
T.~Shiromizu, Y.~Tomikawa, K.~Izumi, and  H.~Yoshino,
PTEP {\bf 2017}, 033E01 (2017);
H.~Yoshino, K.~Izumi, T.~Shiromizu, and  Y.~Tomikawa,
PTEP {\bf 2017}, 063E01 (2017);
D.~V.~Gal'tsov and K.~V.~Kobialko,
Phys. Rev. D {\bf 99}, 084043 (2019).
M.~Siino,
Class. Quantum Grav. {\bf 38}, 025005 (2021);
H.~Yoshino, K.~Izumi, T.~Shiromizu, and Y.~Tomikawa,
PTEP {\bf 2020}, 023E02 (2020);
L.~M.~Cao and Y.~Song,
Eur. Phys. J. C {\bf 81}, 714 (2021);
H.~Yoshino, K.~Izumi, T.~Shiromizu, and  Y.~Tomikawa,
PTEP {\bf 2020}, 053E01 (2020);
K.~Lee, T.~Shiromizu, H.~Yoshino, K.~Izumi, and Y.~Tomikawa,
PTEP {\bf 2020}, 103E03 (2020);
K.~Izumi, Y.~Tomikawa, T.~Shiromizu, and H.~Yoshino,
PTEP, {\bf 2021}, 083E02 (2021);
M.~Siino,
[arXiv:2107.06551 [gr-qc]].

\bibitem{Iyer:2006cn}
  S.~V.~Iyer and A.~O.~Petters,
  Gen.\ Rel.\ Grav.\  {\bf 39}, 1563 (2007).

\bibitem{Tsukamoto:2016qro} 
  N.~Tsukamoto,
  Phys.\ Rev.\ D {\bf 94}, 124001 (2016).

\bibitem{Tsukamoto:2016jzh} 
  N.~Tsukamoto,
  Phys.\ Rev.\ D {\bf 95}, 064035 (2017).

\bibitem{Eiroa:2002mk} 
  E.~F.~Eiroa, G.~E.~Romero, and D.~F.~Torres,
  Phys.\ Rev.\ D {\bf 66}, 024010 (2002).

\bibitem{Eiroa:2003jf} 
  E.~F.~Eiroa and D.~F.~Torres,
  Phys.\ Rev.\ D {\bf 69}, 063004 (2004).

\bibitem{Bozza:2004kq} 
  V.~Bozza and L.~Mancini,
  Astrophys.\ J.\  {\bf 611}, 1045 (2004).

\bibitem{Bozza:2005tg} 
  V.~Bozza, F.~De Luca, G.~Scarpetta, and M.~Sereno,
  Phys.\ Rev.\ D {\bf 72}, 083003 (2005).

\bibitem{Bozza:2006sn}
V.~Bozza and M.~Sereno,
Phys. Rev. D {\bf 73}, 103004 (2006).

\bibitem{Bozza:2006nm} 
  V.~Bozza, F.~De Luca, and G.~Scarpetta,
  Phys.\ Rev.\ D {\bf 74}, 063001 (2006).

\bibitem{Bozza:2007gt} 
V.~Bozza and G.~Scarpetta,
Phys.\ Rev.\ D {\bf 76}, 083008 (2007).

\bibitem{Ishihara:2016sfv} 
  A.~Ishihara, Y.~Suzuki, T.~Ono, and H.~Asada,
  Phys.\ Rev.\ D {\bf 95}, 044017 (2017).

\bibitem{Tsukamoto:2016oca}
N.~Tsukamoto and Y.~Gong,
Phys. Rev. D {\bf 95}, 064034 (2017).

\bibitem{Tsukamoto:2017edq} 
  N.~Tsukamoto,
  Phys.\ Rev.\ D {\bf 95}, 084021 (2017).

\bibitem{Aldi:2016ntn}
G.~F.~Aldi and V.~Bozza,
JCAP {\bf 02}, 033 (2017).

\bibitem{Hsieh:2021scb}
T.~Hsieh, D.~S.~Lee, and C.~Y.~Lin,
Phys.\ Rev.\ D {\bf 103}, 104063 (2021).

\bibitem{Takizawa:2021gdp}
K.~Takizawa and H.~Asada,
Phys. Rev. D {\bf 103}, 104039 (2021).

\bibitem{Tsukamoto:2020bjm}
N.~Tsukamoto,
Phys. Rev. D {\bf 103}, 024033 (2021).

\bibitem{Tsukamoto:2021caq}
N.~Tsukamoto,
Phys. Rev. D {\bf 104}, 064022 (2021).

\bibitem{Aratore:2021usi}
F.~Aratore and V.~Bozza,
JCAP {\bf 10}, 054 (2021).

\bibitem{Chiba:2017nml}
T.~Chiba and M.~Kimura,
PTEP {\bf 2017}, 043E01 (2017).

\bibitem{deVries:2000} 
A. de Vries,
Class. Quantum Grav. {\bf 17}, 123 (2000).

\bibitem{Takahashi:2005hy} 
  R.~Takahashi,
  Publ.\ Astron.\ Soc.\ Jap.\  {\bf 57}, 273 (2005).

\bibitem{Zakharov:2014lqa} 
  A.~F.~Zakharov,
  Phys.\ Rev.\ D {\bf 90}, 062007 (2014).

\bibitem{Akiyama:2019eap}
K.~Akiyama \textit{et al.} [Event Horizon Telescope Collaborations],
Astrophys. J. Lett. \textbf{875}, L6 (2019).

\bibitem{Kocherlakota:2021dcv}
P.~Kocherlakota \textit{et al.} [Event Horizon Telescope Collaboration],
Phys. Rev. D {\bf 103}, 104047 (2021).

\bibitem{Sereno:2003nd} 
  M.~Sereno,
  Phys.\ Rev.\ D {\bf 69}, 023002 (2004).

\bibitem{Bin-Nun:2010exl}
A.~Y.~Bin-Nun,
Phys. Rev. D {\bf 82}, 064009 (2010).

\bibitem{Bin-Nun:2010lws}
A.~Y.~Bin-Nun,
Class. Quant. Grav. {\bf28}, 114003 (2011).

\bibitem{Tsukamoto:2021fsz}
N.~Tsukamoto,
Phys. Rev. D {\bf 104} 124016 (2021).

\bibitem{Holz:2002uf}
  D.~E.~Holz and J.~A.~Wheeler,
  Astrophys.\ J.\  {\bf 578}, 330 (2002).

\bibitem{DePaolis:2003ad}
  F.~De Paolis, G.~Ingrosso, A.~Geralico, and A.~A.~Nucita,
  Astron.\ Astrophys.\  {\bf 409}, 809 (2003).

\bibitem{DePaolis:2004xe}
  F.~De Paolis, A.~Geralico, G.~Ingrosso, A.~A.~Nucita, and A.~Qadir,
  Astron.\ Astrophys.\  {\bf 415}, 1 (2004).

\bibitem{Abdujabbarov:2017pfw}
A.~Abdujabbarov, B.~Ahmedov, N.~Dadhich and F.~Atamurotov,
Phys. Rev. D {\bf 96}, 084017 (2017).

\bibitem{ZamanBabar:2021zuk}
G.~Zaman Babar, F.~Atamurotov, and A.~Zaman Babar,
[arXiv:2104.01340 [gr-qc]].

\bibitem{Bozza:2008ev}
V.~Bozza,
Phys. Rev. D {\bf 78}, 103005 (2008).

\bibitem{Witt:1994}
H.~J.~Witt and S.~Mao, ApJ, {\bf 430}, 505 (1994).

\bibitem{Nemiroff:1994uz}
  R.~J.~Nemiroff and W.~A.~D.~T.~Wickramasinghe,
  Astrophys.\ J.\  {\bf 424}, L21 (1994).

\bibitem{Alcock:1997fi}
  C.~Alcock {\it et al.} [MACHO and GMAN Collaborations],
  Astrophys.\ J.\  {\bf 491}, 436 (1997).

\end{thebibliography}
\end{document}